\documentclass[12pt]{iopart}
\newcommand{\be}{\begin{equation}}
\newcommand{\ee}{\end{equation}}
\newcommand{\bea}{\begin{eqnarray}}
\newcommand{\eea}{\end{eqnarray}}
\newcommand{\frp}[2]{\frac{\partial#1}{\partial#2}}
\newcommand{\dg}{^{\dagger}}
\newcommand{\ex}[1]{{\rm e}^{#1}}
\newcommand{\de}{{\rm d}}
\newcommand{\te}{\tilde{\epsilon}}
\newcommand{\tG}{\tilde G}
\newcommand{\kB}{k_{\rm B}}
\newcommand{\nicht}[1]{ }
\newcommand{\ie}{{\rm i}}
\newcommand{\nn}{\nonumber \\}
\begin{document}

\title{Flow Equations and Normal Ordering}
\author{Elmar K\"ording\footnote{present address: School of Physics and
Astronomy, University of Southampton, SO17 1BJ, UK}~ and Franz Wegner}
\address{Institut f\"ur Theoretische Physik, Universit\"at Heidelberg,
Philosophenweg 19, D-69120 Heidelberg}

\begin{abstract}
In this paper we consider flow-equations where we allow a normal ordering which
is adjusted to the one-particle energy of the Hamiltonian. We show that this
flow converges nearly always to the stable phase. Starting out from the
symmetric Hamiltonian and symmetry-broken normal ordering nearly always yields
symmetry breaking below the critical temperature.
\end{abstract}

\section{Introduction}

In applying flow equations to a Hamiltonian \cite{FW} one typically starts out
from a
Hamiltonian which does not show an explicit symmetry breaking even if the
symmetry of the system will be broken below some temperature.
As in the calculations on the Hubbard model \cite{GW,HW1,HW2,HW3} one can
direct the flow to a form in which molecular-field approximation becomes exact.
This effective Hamiltonian is still not symmetry broken. Only the
molecular-field formalism breaks the symmetry below the critical temperature
$T_c$.

In the fermionic renormalization group flow \cite{S1,S2,ZS,HM,SH} the vertex
functions will at least within
weak-coupling approximations diverge at some length scale below $T_c$, so that
the
approximations become unreliable and one has to resort to other methods in this
regime. Thus it is desirable to have a way to introduce symmetry breaking from
the beginning. Recently Salmhofer, Honerkamp, Metzner, and
Lauscher \cite{SHML} have added a symmetry breaking field to the
Hamiltonian and showed that this additional field leads into the symmetry
broken phase.

For the Hamiltonian flow it is not necessary to add a symmetry breaking term to
the Hamiltonian. Instead it is sufficient to choose a normal-ordering which is
symmetry broken (compare \cite{K}). One can show that the system will nearly
always converge to
the stable state, that is in case of symmetry breaking (that is below the
critical temperature) it runs to a symmetry broken state, whereas if the
symmetric state is stable (above $T_c$) then it will run to the symmetric state.

With this approach the normal ordered form of the Hamiltonian shows an explicit
dependence on the normal ordering. We adjust continuously the normal ordering
to the one-particle contribution of the Hamiltonian and we can show that the
normal ordering has the interesting property to run in the direction of the
stable state. This implies that starting out with a symmetry-broken normal
ordering, then above the critical temperature it will converge to the symmetric
state whereas below criticality it converges to a symmetry broken state.

We start out from the flow equation \cite{FW}
\be
\frac{\de H(l)}{\de l} = [\eta(l),H(l)],
\ee
where $\eta$ is chosen so that the Hamiltonian approaches a diagonal or
block-diagonal form. We write $H$ in the normal-ordered form
\be
H = :H_G:_G
\ee
where $G$ defines the normal ordering
\be
G_{kj} = \langle a_k a_j \rangle_0.
\ee
Here we denote both creation and annihilation operators by $a$.
Then the flow equation reads
\be
\frac{\de H_G}{\de l} = [\eta,H]_G + \frac{\delta H_G}{\delta G} \frp Gl.
\label{flow}
\ee
The second term indicates the change of $H_G$ due to the change of $G$. $G$
itself can be described as the expectation values of corresponding to a
bilinear Hamiltonian $H^0$.
Then the flow equation (\ref{flow}) is supplemented by a second equation for
$H^0$. This equation describes the adaption of $H^0$ to the one-particle energy
of $H$, and will be given later (\ref{tev}).

In the next section we consider the effect of normal ordering on operators.
Then we give the general relation between a bilinear Hamiltonian and the
expectation values in thermal equilibrium. Then we return to the flow equation
and show that almost always the normal ordering will approach a stable state.

\section{Normal Ordering}

The idea behind normal ordering is to subtract expectation values
\be
G_{kj} = \langle a_k a_j\rangle
\ee
from products of operator pairs $a_k$, $a_j$.
More precisely one defines normal ordering which is indicated by two colons by
\bea
:1: &=& 1,\\
:\alpha A(a) + \beta B(a): &=& \alpha :A(a): + \beta :B(a):, \\
a_k :A(a): &=& :a_k A(a): + \sum_j G_{kj} : \frp{}{a_j} A(a):, \label{rec}
\eea
where $\alpha$, $\beta$ are c-numbers. These equations hold for bosons and
fermions. In performing the derivatives for fermions one has to consider that
the operators $a$ anticommute. Iteration of (\ref{rec}) yields
\be
a_{k_1}a_{k_2}...a_{k_m} = :(a_{k_1}+\sum_{l_1}G_{k_1j_1}\frp{}{a_{j_1}})
(a_{k_2}+\sum_{j_2}G_{k_2j_2}\frp{}{a_{j_2}})...a_{k_m}:
\ee
which can also be written
\be
a_{k_1}a_{k_2}...a_{k_m}=:\exp\left(\sum_{kj} G_{kj}
\frp{^2}{a_j^{\rm right}\partial a_k^{\rm left}}\right)
a_{k_1}a_{k_2}...a_{k_m}: \label{prod}
\ee
This is Wick's first theorem \cite{Wick}. The superscripts $^{\rm left}$ and
$^{\rm right}$
indicate that we always pick a pair of factors $a$ and perform the derivative
$\frp{}{a_k}$
on the left factor and the derivative $\frp{}{a_j}$ on the right factor, so
that the factor
$G_{kj}$ depends on the sequence of the operators. The exponential appears in
the equation for the following reason. If we perform the operation 
$G\frp{^2}{a\partial a}$ on $m$ pairs of
factors $a$, then there are due to the permutation symmetry $m!$ contributions.
Therefore in order to obtain the contribution with factor one we have to divide
the $m$-th power of
$G\frp{^2}{a\partial a}$ by $m!$, which yields the exponential. Note that for
fermions the
operators $a$ as well as the derivatives $\frp{}a$ anticommute.

Since in the following we will change the normal-ordering it is appropriate to
indicate to which expectation values $G$ it is performed. Then we obtain quite
general for operators $A$
\be
:A_G:_G = A, \quad
A_G = \exp\left(\sum_{kj} G_{kj} \frp{^2}{a_j^{\rm right}\partial a_k^{\rm
left}}\right) A(a)
\ee
If we dissect
$G_{kj}=Q_{kj}+C_{kj}$, where $C_{kj}=C_{jk}$ for bosons and $C_{kj}=-C_{jk}$
for fermions, then we have
\be
:A_G:_G = :A_Q:_Q, \quad
A_G = \exp\left(\frac 12 \sum_{kj} C_{kj} \frp{^2}{a_j\partial a_k}\right) A_Q.
\ee
and we need no longer distinguish the sequence of the factors. Suppose we have
\be
H = v^{(0)} + \frac 1{2!} \sum_{kj} v^{(1)}_{kj} a_ka_j
+ \frac 1{4!} \sum_{kjmn} v^{(2)}_{kjmn} a_ka_ja_ma_n
\ee
and $v^{(1)}_{kj}$ and $v^{(2)}_{kjmn}$ are completely symmetric for bosons and
antisymmetric for fermions, resp., then the normal-ordered form reads
\bea
H_G &=& v^{(0)}_G + \frac 1{2!} v^{(1)}_{G,kj} a_ka_j
+ \frac 1{4!} v^{(2)}_{G,kjmn} a_ka_ja_ma_n,\\
v^{(0)}_G &=& v^{(0)} +\frac 12 \sum_{kj}v^{(1)}_{kj}G_{kj}
+\frac 18 \sum_{kjmn} v^{(2)}_{kjmn} G_{kj}G_{mn}, \\
v^{(1)}_{G,kj} &=& v^{(1)}_{kj} + \frac 12 \sum_{mn} v^{(2)}_{kjmn} G_{mn},\\
v^{(2)}_{G,kjmn} &=& v^{(2)}_{kjmn}.
\eea
Expressing $A$ and $B$ by $A_G$ and $B_G$ by means of eq. (\ref{prod}) and then
transforming back to the normal-ordering we obtain for the product of two
operators 
\be
:A_G(a):_G \,\, :B_G(a):_G = :\exp\left(\sum_{kj} G_{kj} \frp{^2}{b_j\partial
a_k}\right)
A_G(a) B_G(b):|_{G,b=a}.
\ee

\nicht{\subsection{Continuous Symmetry}

Suppose the Hamiltonian commutes with a one-particle Operator $L$,
\be
L:= l^{(0)} + \frac 12 l_{km} :a_ka_m:.
\ee
Under a change of the normal-order only the constant $l^{(0)}$ may change,
whereas the coefficients $l_{km}$ are independent of the normal order. Then
$[L,H]=0$ implies
\be
{[}L,H]= -(G_{kj}+G_{jk}) l_{km} :a_m \frp H{a_j}:
-\frac 12 (G_{kj}G_{mn}-G_{jk}G_{nm}) l_{km} : \frp{^2H}{a_n\partial a_j}
: = 0.
\ee
If we use that
\be
G_{kj}+G_{jk} = \langle a_ka_j+a_ja_k \rangle
= \langle a\dg_{k^*}a_j + a_ja\dg_{k^*} \rangle = \delta_{k^*,j}
\ee
then $H$ obeys
\be
{[}L,H]= l_{km} :\left(a_m\frp H{a_{k^*}} 
+ \frac 12 (G_{mj}-G_{jm}) \frp{^2H}{a_{k^*}\partial a_j}\right): = 0.
\ee
} % nicht

\section{Expectation values for $H^0$}

In the following we will consider only fermions.
A conventional way to introduce normal ordering is to use a Hamiltonian
\be
H^0 = \frac 12 \te_{kj} a_{k^*}a_j
\ee
where we use the notation
\be
a\dg_k = a_{k^*}, \quad a_k = a\dg_{k^*}.
\ee
Considering $a$ a column vector and $a\dg$ a row vector we may write
\bea
\te_{kj}a_{k^*}a_j &=& \left(\begin{array}{cc}a\dg & a^T \end{array} \right)
\te \left(\begin{array}c a \\ a^* \end{array}\right)
= \left(\begin{array}{cc}a\dg & a^T \end{array} \right)
\left(\begin{array}{cc} A & B \\ B\dg & -A^T \end{array}\right)
\left(\begin{array}c a \\ a^* \end{array}\right) \nn
&& \quad A\dg = A, \quad B^T=-B.
\eea
Thus $\te$ has the properties
\be
\te\dg=\te, \quad \tau \te \tau = - \te^T, \quad
\tau=\left(\begin{array}{cc} 0 & 1 \\ 1 & 0 \end{array} \right)
\ee
$\te$ can be diagonalized with diagonal matrix elements $\epsilon_k$ and
$-\epsilon_k$. (This diagonalization is performed by a canonical transformation
which is isomorphic to a real orthogonal transformation. This can be easily
seen if one introduces the hermitean linear combinations
$q_k=\frac{a\dg_k+a_k}{\sqrt 2}$ and
$p_k=\frac{\ie(a\dg_k-a_k)}{\sqrt 2}$.)

In thermal equilibrium one obtains for diagonal $\te$, i.e.
$\te_{kj}=\delta_{kj}\epsilon_k$
\be
\langle a\dg_k a_k \rangle = \frac 1{\ex{\beta\epsilon_k}+1}, \quad
\langle a_k a\dg_k \rangle = \frac 1{\ex{-\beta\epsilon_k}+1}
\ee
so that even when $\te$ is not diagonal
\be
\tG = \frac 1{\ex{\beta\te}+1}
= \frac 12 - \frac 12 \tanh(\frac{\beta\te}2)
\ee
holds with $\tG_{kj}=G_{k^*j}$. A variation of $\te$ yields
\be
\delta\tilde G= -\frac 1{\ex{\beta\te}+1} \,\delta\ex{\beta\te}\,
\frac 1{\ex{\beta\te}+1}
=-\frac 1{\ex{\beta\te}+1} \int_0^{\beta}\de\tau \ex{\tau\te} \,\delta\te\,
\ex{(\beta-\tau)\te} \frac 1{\ex{\beta\te}+1}.
\ee
Thus we may write
\be
\delta\tilde G_{kj} = - \Gamma_{kj,pq} \,\delta\te_{pq} \label{rel}
\ee
with
\be
\Gamma_{kj,pq} = \int_0^{\beta} \de\tau \left(\frac 1{\ex{\beta\te}+1}
\ex{\tau\te} \right)_{kp}
\left( \ex{(\beta-\tau)\te} \frac 1{\ex{\beta\te}+1} \right)_{qj}.
\ee

\section{Free Energy and Stability}

It is well known, that for a given Hamiltonian $H$ and temperature $T$ the free
energy assumes its minimum for the corresponding statistical operator
$\rho=\ex{-\beta H}/Z$. Thus one often determines approximately the free energy
for
the statistical operator $\rho^0=\ex{-\beta H^0}/Z^0$ with $H^0$ bilinear in
the operators $a$ and determines $\te$ so that the corresponding free energy
$F^0$ becomes minimal. One obtains
\be
F^0 = E - TS, \quad E= v^{(0)}_G, \quad
S=-\frac{\kB}2 {\rm tr}(\tG \ln\tG + (1-\tG) \ln(1-\tG))
\ee
We use that $\langle A(a) \rangle^0=\left.A_G\right|_{a=0}$.
We will vary this expression. By means of
\bea
A_{G+\delta G} &=& A_G+\frac 12 \sum_{kj} \delta G_{kj} \frp{^2}{a_j\partial
a_k} A_G \\
&+& \frac 18 \sum_{kjmn} \delta G_{kj}\delta G_{mn}
\frp{^4}{a_j\partial a_k\partial a_n \partial a_m} A_G + O(\delta G^3)
\eea
we obtain
\be
E=v^{(0)}_G + \frac 12 v^{(1)}_{G,k^*j} \delta\tG_{kj}
+\frac 18 v^{(2)}_{G,k^*jm^*n} \delta\tG_{kj} \delta\tG_{mn} + ...
\ee
A variation of $S$ yields in first order in $\delta\tG$
\be
\delta S=-\frac{\kB}2 \ln\left(\frac{\tG}{1-\tG}\right)_{jk} \delta\tG_{kj}
\ee
Therefore we obtain in first order in $\delta\tG$
\bea
\delta F^0 &=& \frac 12 \left( v^{(1)}_{G,k^*j}
+ \kB T \ln\left(\frac{\tG}{1-\tG}\right)_{jk} \right) \delta\tG_{kj} \nn
&=& \frac 12 ( v^{(1)}_{G,k^*j}-\te_{kj}) \delta\tG_{kj}. \label{gleich}
\eea
In order that $F^0$ is an extremum one has to choose
$\te_{kj}=v^{(1)}_{G,k^*j}$, that is the one-particle contribution of our
Hamiltonian $H_G$ has to agree with $H^0$.
In second order in $\delta\tG$ we obtain
\be
\delta F^0 = \frac 18 v^{(2)}_{G,k^*jm^*n} \delta\tG_{kj} \delta\tG_{mn}
-\frac 14 \delta\te_{kj} \delta\tG_{kj},
\ee
where $\delta\te$ is given in terms of $\delta\tG$ from (\ref{rel})
\be
\delta\te_{kj} = -(\Gamma^{-1})_{kj,mn} \delta\tG_{mn}.
\ee
which yields
\be
\delta F^0 = \frac 14 (\frac 12 v^{(2)}_{G,k^*jm^*n} + (\Gamma^{-1})_{kj,mn} )
\delta\tG_{kj} \delta\tG_{mn}. \label{F0}
\ee
Only if this expression is positive definite then it corresponds to a stable
solution.
We note that $\Gamma$ is positive definite and therefore $\Gamma^{-1}$ exists
and is positive definite, too. This can be seen if we switch to the basis in
which $\te$ is diagonal, $\te_{kj}=\delta_{kj}\epsilon_k$. Then
\bea
\Gamma_{kj,pq} &=& \delta_{kp} \delta_{jq} \Gamma^{\rm d}_{kj}, \\
\Gamma^{\rm d}_{kj} &=& \frac{\ex{\beta\epsilon_k}-\ex{\beta\epsilon_j}}
{(\ex{\beta\epsilon_k}+1)(\epsilon_k-\epsilon_j)(\ex{\beta\epsilon_j}+1)}>0.
\eea
Thus $\Gamma$ is diagonal with positive matrix elements along the diagonal.

These expressions are closely connected to the response function in random
phase approximation. Suppose one adds a perturbation $\delta v^{(1\,\rm ext)}$
to the Hamiltonian $H$ which produces a change $\delta\tG^{\rm ext}$ in the
expectation values
$\langle a_k a_j \rangle$ then we have from eqs. (\ref{gleich},\ref{F0})
\be
\frac 12 \left(\delta v^{(1,\,\rm ext)}_{G,k^*j}
+(\frac 12 v^{(2)}_{G,k^*jm^*n}+(\Gamma^{-1})_{kj,mn}) \delta \tG_{mn} \right)
\delta\tG^{\rm ext}_{kj}=0.
\ee
If we now relate $\delta\tG^{\rm ext}$ to a perturbation $\delta v^{(1,\rm
eff)}$ which would have the same effect without the interaction $v^{(2)}$
according to eq. (\ref{rel})
\be
\delta\tG^{\rm eff}_{mn} = - \Gamma_{mn,pq} \,\delta v^{(1,\rm eff)}_{pq}
\ee
then we obtain
\be
\delta v^{(1,\,\rm ext)}_{G,k^*j} = (\frac 12 v^{(2)}_{G,k^*jm^*n}
\Gamma_{mn,pq} +\delta_{kp}\delta_{jq}) \delta v^{(1,\rm eff)}_{p^*q}
\ee
This factor $\frac 12 v^{(2)}\Gamma +1$ enters for example in the Lindhard
expression for the static dielectric constant. One has only to introduce
$v^{(2)}\propto e^2/q^2$ and
\be
\Gamma^{\rm d}_{kj} = \frac{f(\epsilon_j)-f(\epsilon_k)}{\epsilon_k-\epsilon_j}
\ee
with the Fermi function $f(\epsilon)$. The dielectric constant has to be
positive for a stable system.

\section{Flow Equations}

Now we have to introduce our flow equations. We have two contributions to the
change
of $H_G$, one from the generator of the flow,
\be
{[}\eta,H]_G= g^{(0)} + \frac 12 g^{(1)}_{kj} a_ka_j + ...
\ee
and one from the change of the normal-ordering which yields
\be
\frac 12 \frp{G_{kj}}l \frp{^2}{a_j\partial a_k}H =
-\frac 12 \Gamma_{kj,mn} \frp{\te_{mn}}l \left(v^{(1)}_{k^*j}
+ \frac 12 v^{(2)}_{k^*jp^*q}a_{p^*}a_{q}+...\right)
\ee
Thus we obtain the change of $v^{(1)}$
\be
\frp{v^{(1)}_{k^*j}}l = g^{(1)}_{k^*j} - \frac 12 \Gamma_{pq,mn}
v^{(2)}_{p^*qk^*j}
\frp{\te_{mn}}l
\ee
for the one-particle energy. On the other hand we wish to adapt $\te$
so that it approaches $v^{(1)}$. Therefore we introduce a flow equation for
$\te$
\be
\frp{\te_{kj}}l = \gamma (v^{(1)}_{k^*j} - \te_{kj}) \label{tev}
\ee
with some positive constant $\gamma$. Evidently for fixed $v^{(1)}$ the energy
$\te$  approaches exponentially $v^{(1)}$.
From both equations we obtain
\be
\frp{(v^{(1)}_{k^*j}-\te_{kj})}l = g^{(1)}_{k^*j}
-\gamma(\delta_{km}\delta_{jn}+\frac 12 \Gamma_{pq,mn} v^{(2)}_{p^*qk^*j})
(v^{(1)}_{m^*n}-\te_{mn})
\ee
Thus as $g^{(1)}$ from the generator of the flow decreases also $v^{(1)}-\te$
will decrease
provided the kernel $1+\frac 12 \Gamma v^{(2)}$ is positive definite. Otherwise
the difference $v^{(1)}-\te$ will nearly always increase exponentially. The
condition that this kernel is positive definite is equivalent to the stability
obtained from $F^0$ in eq. (\ref{F0}) above, since the kernel for $F^0$ differs
only by the factor $\Gamma$, which itself is positive definite.

\end{document}